# HAPI-FHIR Server Implementation to Enhancing Interoperability among Primary Care Health Information Systems in Sri Lanka: Review of the Technical Use Case

Prabath Jayathissa[1], Roshan Hewapathrana[2]
[1]Postgraduate Institute of Medicine, University of Colombo, Sri Lanka
[2]Faculty of Medicine, University of Colombo, Sri Lanka

**Abstract.** This comprehensive review underscores the paramount importance of interoperability within the digital health landscape, emphasizing the necessity for a standardized framework to facilitate effective communication among healthcare professionals and institutions. The primary focus of this discourse centres on implementing a Fast Healthcare Interoperability Resources (FHIR) server, recognised as a pivotal solution addressing technical, semantic, and process interoperability failures. This standardised framework ensures uniformity and facilitates efficient communication and real-time data access within Primary Care Health Information Systems (HIS). The adaptability and scalability inherent in FHIR play a critical role in supporting the dynamic needs of healthcare systems, fostering interoperability, and enabling integration across diverse components. The narrative delves into the complexities of patient data management, accentuating the pivotal role of semantic interoperability in ensuring the seamless continuation of patient care. The transition from paper-based documentation to repository storage necessitates effective data retrieval through clinical correlation, emphasising presenting health data in a manner aligned with clinical findings—an innovative concept introduced as a health-aware presentation. Integrating FHIR standards amplifies these efforts, enriching multiple pathways for data search and retrieval. This interconnectedness not only fosters efficient interoperability within healthcare institutions but also facilitates a comprehensive approach to accessing health data across diverse organizations. The FHIR server implementation project, guided by the principles of the ADR method, systematically addresses challenges associated with patient identity criteria, biometrics, and data security, demonstrating a steadfast commitment to inclusive and patientcentric care. The detailed exploration of the development phases of the FHIR server implementation accentuates the significance of architecture design, API integration, and security measures. The concluding stages underscore a forward-looking approach, incorporating HHIMS Synthetic Dataset testing for future utilization. Ultimately, positioning the integration of an FHIR server in Primary Care HIS as a transformative step, this abstract envisions the fostering of a dynamic and responsive healthcare information environment harmonizing with the evolving landscape of digital health.

**Introduction**

**Primary Care Health Information Systems in Sri Lankan**

In Sri Lanka, the government health care system encompasses various service deliveries, including administration, training, curative care, public health, and resource management programs. The curative care component involves hospitals categorized based on resource allocation and availability. Teaching hospitals, provincial general and district general hospitals, special care units, and rehabilitation units fall under the line ministry, which serves as the core of the health ministry. Other hospitals are affiliated with provincial councils under the provincial health ministry (Ministry of Health, Nutrition and Indigenous Medicine - Sri Lanka, 2017).

The curative sector operates on a referral system, directing patients from primary to tertiary care institutions. This system aims to ensure the sustainability of healthcare delivery, minimizing delays for both paediatric and adult populations. The public sector provides 95% of inpatient care, while the private sector is responsible for the remaining 5% However, the referral system doesn't prevent direct engagement with private practices or admission to private hospitals (Ministry of Health - Sri Lanka, 2018). Primary Care Health Information Systems in Sri Lanka play a pivotal role in the country's healthcare landscape, serving as the foundation for the delivery of essential and preventive healthcare services. These systems encompass a range of technological tools and platforms designed to manage, store, and exchange patient information within the primary care setting. Understanding the dynamics of these information systems is crucial for effective healthcare delivery and the overarching goal of improving population health in Sri Lanka (Perera et al., 2019).

Primary care health information systems in Sri Lanka play a crucial role in modern healthcare delivery. These systems, primarily based on electronic health records (EHRs), centralize patient information, including medical history, treatment plans, and diagnostic data. The digitalization of records improves accessibility and facilitates coordinated care among healthcare providers (Al-Sahan & Saddik, 2016). Efficient appointment scheduling and workflow management are integral components, ensuring organized and patient-friendly healthcare experiences. Health data interoperability is emphasized, allowing seamless exchange among different facilities for comprehensive patient care. Integration of telemedicine solutions enables remote consultations and enhances accessibility, particularly in remote or underserved areas (Yuksel et al., 2016).

These systems contribute to public health initiatives by supporting surveillance and reporting. They monitor disease trends, identify outbreaks, and aid in implementing preventive measures. Robust data security measures prioritize patient confidentiality and compliance with national and international standards. Decision support systems enhance clinical decision-making, providing alerts for medication interactions, reminders for preventive care, and access to evidence-based guidelines. Understanding these functionalities and addressing challenges is crucial for policymakers, healthcare providers, and technology developers as primary care health information systems continue to shape the future of healthcare delivery in Sri Lanka (Jayathissa, Vajira, & Roshan, 2017a).

The concept of universal healthcare, rooted in history, led to global initiatives like WHO. In Sri Lanka, Primary Healthcare Reform addresses challenges, emphasizing the Primary Care System Strengthening Project (PSSP). PSSP aims to improve primary healthcare utilization, focusing on health information system interoperability. The Cluster Master Patient Index is proposed for sharing records. The project includes implementing a Health Information Management System and a Personal Health Records system (Ministry of Health - Sri Lanka, 2018). The Sri Lanka Essential Service Package ensures citizens receive services, emphasizing preventive care and strengthening referrals. The empanelment of citizens into clusters ensures a concise health record, promoting continuity of care. Overall, these initiatives drive universal health coverage in Sri Lanka.

According to Sri Lanka's constitution, anyone in the country can receive treatment at any hospital, with free healthcare services accessible nationwide. However, patient bypassing primary care institutes creates unnecessary resource competition at secondary levels. The Essential Service Package (ESP) acts as a gatekeeper for referrals, aiming to strengthen Primary Medical Care Institutions (PMCI) credibility and attract patients (Ministry of Health - Sri Lanka, 2018). A quality referral system requires standardized processes, appropriate education for healthcare professionals, and supervision for adherence. Effective referrals involve prompt and respectful patient treatment at PMCI, and decision-making is based on analysed clinical signs. Despite patient autonomy, clear referral letters facilitate seamless communication for timely attention at the receiving institute (Jayathissa & Hewapathirana, 2023b).

**Hapi-FHIR Server**

Although the health information systems were set up like mushrooms, came to this research with a strong understanding of the effects that patients may face due to the lack of interactions between health information systems (Jayathissa & Hewapathirana, 2023c; Jayathissa, Vajira, & Roshan, 2017a). To improve the continuation of care, universal care coverage should be established. To achieve universal health care coverage, the lack of interaction between health information systems should be overcome with health information system interoperability. Then the health records are shared among the primary and secondary care. The primary healthcare strengthening project is the first attempt to achieve universal healthcare through health information system interoperability. To bridge the gap, the perfect solution could be the interoperability service (Jayathissa & Hewapathirana, 2023a).

With the establishment of the Essential Service Package (ESP) in 2018, primary health care reached a unique level. With the introduction of ESP, patients are empanelled into clusters, which are designed to provide the best possible care for patients. Still, it is felt that the primary health information systems are not interoperable with each other to ensure that services are properly accessed (Jayathissa et al., 2018).

In the dynamic landscape of healthcare, achieving interoperability among primary care health information systems is paramount to enhance the overall quality of patient care and streamline healthcare processes. This document explores the implementation of the Health Level Seven International's (HL7) Fast Healthcare Interoperability Resources (FHIR) standard within the context of Sri Lanka's primary care health systems. Specifically, it delves into the HAPIFHIR Server Implementation, evaluating its effectiveness in fostering interoperability and facilitating seamless data exchange among diverse health information systems. Through an in-depth review of the technical use case, this study aims to provide valuable insights into the practical applications, challenges, and potential benefits of adopting HAPIFHIR within the unique healthcare ecosystem of Sri Lanka. As we navigate this exploration, we seek to unravel the transformative potential of this implementation, ultimately contributing to the advancement of interoperable and patient-centric healthcare solutions in the region (Jayathissa, Vajira, & Roshan, 2017a).

The evolution of healthcare systems globally has underscored the critical importance of interoperability, as it empowers healthcare providers with the ability to access, share, and utilize patient information seamlessly. In the context of Sri Lanka's primary care health information systems, the HAPI-FHIR Server Implementation emerges as a promising avenue to address interoperability challenges. This document embarks on a comprehensive examination of the technical use case of the HAPI-FHIR Server Implementation in the Sri Lankan healthcare landscape. The review not only scrutinizes the intricacies of the implementation but also aims to shed light on the contextual nuances and specific requirements of primary care health systems in Sri Lanka. By investigating the alignment of HAPI-FHIR with the unique challenges faced by the Sri Lankan healthcare ecosystem, this study seeks to provide actionable insights that can inform decision-makers, developers, and stakeholders. The discussion on interoperability in Sri Lanka's health information systems began in 2017, with the proposal to interconnect these systems. The Master Patient Index (MPI) emerged as a suitable tool for serving as the integration layer for this task. The MPI, with its various functionalities, particularly its role in interoperability via messaging, was identified as a key element in this effort. Development efforts were underway to achieve interoperability using the microservice architecture mentioned in the 2017 paper (Jayathissa & Hewapathirana, 2023a; Jayathissa, Dissanayake, & Hewapathirana, 2018; Jayathissa et al., 2023).

Patient matching is a critical function of the MPI, and the use of probabilistic matching helps prevent duplications. The development of management capacity and improvement in electronic communication were identified as factors that could streamline and standardize electronic communication in health information systems (Jayathissa & Hewapathirana, 2022; Jayathissa, Vajira, & Roshan, 2017a). As technology, particularly the internet and information communication technologies, continues to advance, the adoption of e-health initiatives is expected to transform traditional healthcare systems globally. This transformation aims to reduce medical errors, enhance healthcare quality, minimize costs, and empower clients to actively engage in their healthcare decisions.

For successful interoperability among health information systems, effective communication channels are crucial. The MPI is recognized globally as a primary tool for interconnecting health information systems. Implementing the MPI is seen as instrumental in achieving interoperability among health records in the context of primary healthcare reform. The MPI is envisioned to play a pivotal role in achieving cluster-based referral governance through system interoperability, making it a key element in the quest for interoperability in healthcare information systems (HL7, 2014; Marceglia et al., 2015). As we navigate through the technical intricacies of the HAPI-FHIR Server Implementation, it becomes imperative to explore how this framework facilitates data exchange, supports standardized data formats, and promotes a patientcentric approach to healthcare. Additionally, the document will delve into potential challenges and considerations associated with the integration of HAPI-FHIR within the Sri Lankan context (HL7, 2017; Hussain, Langer, & Kohli, 2018; University Health Network, 2018).

In essence, this review not only serves as a technical evaluation but also as a roadmap for enhancing interoperability among primary care health information systems in Sri Lanka. By critically assessing the HAPI-FHIR Server Implementation and its alignment with the unique needs of the region, we aspire to contribute to the ongoing discourse on leveraging technology for improved healthcare delivery and patient outcomes (Hussain, Langer, & Kohli, 2018).

Beyond the technical intricacies, this document also explores the broader implications of implementing HAPI-FHIR in the Sri Lankan healthcare landscape. It delves into the potential impact on data security, privacy considerations, and the overall governance framework required for sustaining a robust interoperable ecosystem (HL7, 2014). Furthermore, the review seeks to assess the scalability of the HAPI-FHIR Server Implementation, considering the diverse and evolving needs of primary care health systems in Sri Lanka. Understanding how this solution adapts to changing healthcare requirements and accommodates future advancements is crucial for its long-term viability and success (Hussain, Langer, & Kohli, 2018; Pimenta et al., 2023).

An integral part of this exploration involves examining the collaborative potential of HAPI-FHIR, both within and beyond national borders. Given the increasing importance of cross-border healthcare collaborations and the global nature of health challenges, understanding how HAPI-FHIR supports interoperability on an international scale becomes pivotal. Additionally, the document aims to highlight any success stories or best practices emerging from the implementation of HAPI-FHIR in primary care settings. Real-world examples of improved patient outcomes, streamlined workflows, and enhanced data-driven decision-making can serve as compelling evidence for the efficacy of this implementation, providing valuable insights for other regions contemplating similar endeavours (Soares, Essaid, & Kahn, 2023).

In conclusion, this review aspires to offer a comprehensive understanding of the HAPI-FHIR Server Implementation's impact on interoperability among primary care health information systems in Sri Lanka. By navigating through technical nuances, addressing contextual challenges, and envisioning the broader implications, this document seeks to contribute to the ongoing dialogue on advancing healthcare through innovative and interoperable solutions in the Sri Lankan context.

**Methods**
The research method involves a systematic approach aimed at achieving a comprehensive understanding and successful implementation of the intended objectives. To begin, the research objective is clearly defined: to implement the HAPIFHIR server for enhancing interoperability among Primary Care Health Information Systems (HIS). This serves as the guiding principle throughout the research process.

In the pursuit of enhancing interoperability through HAPI-FHIR server implementation, conducting a comprehensive literature search is imperative to gather insights from relevant research articles. Leveraging academic databases such as PubMed, IEEE Xplore, ScienceDirect, and Google Scholar can serve as a starting point. Employing keywords like "HAPI-FHIR," "FHIR Server Implementation," and "Interoperability" with Boolean operators helps refine the search. Filtering results by publication date ensures access to the latest information, while review articles offer comprehensive overviews. Examining citations in relevant articles and exploring journals, conferences, government websites, and professional associations in healthcare informatics can unveil valuable sources. Collaboration with librarians can further enhance the search strategy. Critical evaluation of the identified articles for relevance and quality is crucial for obtaining meaningful insights into the topic. As the search query ("HAPI-FHIR" OR "Fast Healthcare Interoperability Resources") AND ("Server Implementation" OR "Implementation Strategies") AND ("Interoperability" OR "Health Information Exchange") were been used.

A crucial step involves conducting an extensive literature review focusing on FHIR standards, interoperability in healthcare, and existing implementations. This review aims to identify challenges and gaps within current systems, providing a foundation for the subsequent research phases. Defining the system requirements comes next, involving the specification of both functional and non-functional requirements based on the identified gaps and challenges. This step ensures a clear roadmap for the implementation process. The selection of representative Primary Care HIS for integration follows, considering factors such as user base, data complexity, and system architecture. This step is pivotal in ensuring the practical relevance and applicability of the research outcomes.

The actual implementation phase involves configuring the HAPI-FHIR server to meet the defined requirements and ensuring compatibility with the selected Primary Care HIS. Additionally, strategies for data mapping and transformation are developed to address issues related to data types, formats, and standards. Security measures are implemented to ensure the confidentiality and integrity of patient data during interoperability, aligning with healthcare data protection standards. Rigorous testing, including unit tests, integration tests, and end-to-end tests, is then conducted to validate the functionality and interoperability of the implemented HAPI-FHIR server. User feedback is collected from healthcare professionals and IT staff to assess the usability and effectiveness of the HAPI-FHIR implementation. Performance evaluation follows, focusing on aspects such as response times, data throughput, and system resource utilization. A scalability assessment is conducted to evaluate the system's capability to accommodate an increasing number of users and data volume. Subsequently, a cost-benefit analysis is performed to assess the HAPI-FHIR implementation's economic feasibility compared to the enhanced interoperability (Pimenta et al., 2023; Soares, Essaid, & Kahn, 2023; Yuksel et al., 2016).

Thorough documentation of the entire implementation process, including configurations, codebases, and integration protocols, is undertaken. Ethical considerations related to patient data privacy, consent, and compliance with healthcare regulations are addressed throughout the research process. In conclusion, the research findings are summarized, conclusions are drawn regarding the effectiveness of the HAPI-FHIR implementation, and recommendations are provided for future improvements. This comprehensive research method ensures a systematic and thorough exploration of the HAPI-FHIR server's role in enhancing interoperability among Primary Care Health Information Systems.

For information system development has two basic missions to complete to achieve the successful product and implementation. First mission is contribution to the theoretical aspect of the research and second aspect is how to show the problems of the system using the current practices. For system development and implementation, the action research (AR) and the design research (DR) methods being used. In the design research, the new design is developed and the action research solves an immediate organizational problem. To make a product as successful one researcher has to put the theory into practice in a practical way. The design research and the action research work in a similar way. Designing the master patient index is a DR and the implementation in the health care setting is an AR. ADR method "IT artifact is built and evaluate in an organizational setting using a general prescriptive design". The research project is a qualitative research. I used action design research (ADR) as the method for HAPI-FHIR Server Implementation (Jayathissa, Vajira, & Roshan, 2017b).

ADR method "IT artifact is built and evaluate in an organizational setting using a general prescriptive design". This deals with 3 main areas of the IT artifact development. 1st problem formulation process using the design research (DR) method .2nd Evaluation using DR method. 3rd and final are the artifact building. As according to the ADR method research project is carried out in 3 phases. In the phase 1 problem formulation done. Phase 2 building, intervention and evaluation process for the MPI Architecture. At the final phase developing the system after formalization of what learn from the previous phases. In the phase 1 data collection using focus groups, discussion and analysis of its result will be done in order to identify data elements and user requirements of MPI prototype. Phase 2 involves in identify the suitable API model for the proposed MPI. Phase 3 is developing a prototype MPI using the identified user requirements.

In the phase 1 did the problem formulation. During this phase practical, inspired research principal used. In this phase viewing the field problem and knowledge gathering and creating opportunities is the main target. Opportunity is the intersection of technical and organizational domains. Action design researcher will get the knowledge to apply to set of problems to solve the problem. This was carried out using Qualitative research design. Formative evaluation be applied to the qualitative research design. The research did on the staff members who are involved in the use of the HHIMS/HIMS health information system in the relevant respective hospitals (Base Hospital Panadura, Base Hospital Horana, Base Hospital Homagama, Base Hospital Avissawella, District hospital Dompe, Cancer Institute Maharagama) of two districts (Colombo and Kalutara). The sampling method is Convenient sampling method is chosen mainly because of the limited resources and time. Study instruments Semi-Structured interviews for HHIMS/HIMS users. Focus group discussions and interviews with, HHIMS/HIMS users and Software Development group team. Interviews and Focus group discussion (FGD) conducted by the principal investigator with the help of supervisors. Data recording was done correctly. Randomly selection of participants/groups for FGD was done. FGD was done following an unstructured interview at the beginning, and then semi-structured questions was created to ask later by avoiding the unimportant issues arrived at the earlier conversation.

After taking permission from the relevant authorities the principal investigator introduced himself to the staff at the above relevant respective hospital of Colombo and Kalutara district and explain the research and the methodology to the eligible participants of the research. Informed consent was obtained from the eligible participants after explaining the full investigation procedure including benefits and risks. An adequate time allowed for them to clarify any query or further questions. Informing them that they can withdraw from the study at any time without giving any reason and no further correspondence made ensures voluntary participation.

Data from interviews analysed using qualitative methods to find relevant concepts related to the implementation. Collected ideas recorded using computer software. These recoded data used to identify themes, Concepts, requirements of the users. Research did not involve collecting personal identification information of participants during data collection. Therefore, privacy and confidentiality of the participants are preserved. Ethical approval has detained from the ethics review committee PGIM, University of Colombo. Before collecting data from health institutions necessary permission gained from the head of the institutes. Informed written consent taken from the participants. Privacy, security, and confidentiality of data assured. The purpose of the research and publication would not involve using any personally identifiable data of anybody or there is no possible access the system or its data by third parties because all data gathering would be done by researcher's personal computer.

Phase 2: During the implementation phase of the HAPI FHIR server, the focus shifted towards the design cycle of the IT artifact. This involved the creation of a design architecture integral to the development process. Collaborative efforts from domain experts were employed to formulate a plan for the architectural design. The development adhered to three fundamental principles, initiating with the construction of the architectural design. Subsequently, interventions were introduced to assess the suitability of the design. The evaluation phase, conducted at the conclusion, leveraged a literature review of existing systems as a tool for comprehensive assessment. In the initial design phase, the research team meticulously crafted the Alpha-version of the HAPI FHIR server architecture. Following a comprehensive literature review and extensive discussions with the development team, the beta version of the design was meticulously formulated.

Advancing into Phase 3, the Software Development Life Cycle (SDLC) commenced, embracing the Scrum software development practice for the effective implementation of the HAPI FHIR server. The agile methodology of Scrum ensured the delivery of software program deliverables at the conclusion of each software development phase, providing transparency for the researcher and delivering incremental progress every two weeks. The Scrum procedure in Phase 3 encompassed crucial steps. A comprehensive work list was established, driven by the researcher's features derived from the Phase 1 requirement document. Tasks were meticulously estimated, prioritized, and organized into sprints—two-week periods dedicated to task delivery. Sprint meetings discussed tasks at the beginning, and deliverables were provided at the end, accompanied by plans for subsequent sprints.

Implementation steps for the HAPI FHIR server involved configuring the server based on the designed architecture, developing APIs, and defining core interactions and message types between the electronic health record and the HAPI FHIR server. This phase also witnessed the creation of a mechanism for incorporating a unique identification number into the HAPI FHIR server, enhancing its functionality. The agile approach was wholeheartedly embraced by the software development team at ICTA, ensuring continuous and incremental progress. A designated software team leader took charge, and the scrum procedure was initiated after the full requirement document was meticulously finalized. Testing of the message architecture involved using the prototype HAPI FHIR server and electronic health records, a process meticulously conducted by the development team. Throughout the development journey, the team adhered to a stringent schedule, with separate teams dedicated to software development, testing, and security. The iterative process included development, rigorous testing, and the implementation of robust security measures at each software deliverable, ensuring a systematic, secure, and successful implementation of the HAPI FHIR server.

**Results**

In the initial design phase, the research team crafted the Alpha-version of the HAPI FHIR server architecture, envisioning a comprehensive FHIR playground using the Docker environment. Establishing a HAPI FHIR server within a Dockerized environment involves encapsulating the server in a Docker container and configuring it for seamless execution. This process commences by composing a Dockerfile, specifying instructions such as utilizing a Java base image, setting the working directory, and integrating the HAPI FHIR server JAR file into the container. Subsequently, the Docker image is built, assigning a relevant tag. Upon successful image construction, the Docker container is executed, mapping the local machine's port 8080 to the container's port 8080. The operational HAPI FHIR server can then be accessed through a web browser at http://localhost:8080/baseDstu3, with adjustments made for the specific FHIR version. Custom configurations, if needed, involve incorporating volumes or transferring configuration files into the container. This may include copying a customized hapi-config.properties file and adjusting the CMD instruction in the Dockerfile accordingly. Adherence to the HAPI FHIR server documentation for configuration details and version specific considerations is crucial, and compliance with licensing agreements is imperative in a production environment.

With experience of the Docker implementation 2nd stage started with implementation involved integrating essential components and tools, including specific prerequisites such as the Java Development Kit (JDK), Java Authentication and Authorization Service (JAS), Apache Maven, Rubz, and Budler. The subsequent steps in the HAPI FHIR server implementation encompassed an extensive literature review on FHIR standards, interoperability, and existing implementations. Functional and non-functional requirements were defined based on identified gaps and challenges. A representative Primary Care Health Information System (HIS) was selected for integration, considering user base and system architecture. The HAPI FHIR server was then implemented and configured to ensure compatibility with the chosen Primary Care HIS.

Data mapping strategies were developed to address issues related to data types, formats, and standards, while security protocols were implemented to ensure data confidentiality and integrity during interoperability. Rigorous testing, including unit tests, integration tests, and end-to-end tests, was conducted. User feedback was collected from healthcare professionals and IT staff to assess usability. The server's performance was evaluated in terms of response times, data throughput, and resource utilization. Scalability was assessed to accommodate an increasing number of users and data volume. A cost-benefit analysis was conducted to evaluate economic feasibility, and the entire implementation process, including configurations and integration protocols, was thoroughly documented. Ethical considerations related to patient data privacy, consent, and healthcare regulations were addressed. The findings were summarized, conclusions were drawn on the effectiveness of the HAPI FHIR implementation, and recommendations were provided for future improvements (Jayathissa, Vajira, & Roshan, 2017b).

Simultaneously, the team embarked on building a customized FHIR playground. Prerequisites, including JDK, JAS, Apache Maven, Rubz, and Budler, were ensured. The HAPI FHIR server was compiled and installed, incorporating necessary tools and ensuring proper configuration. This resulted in the establishment of a versatile FHIR playground, allowing hands-on exploration of FHIR resources and services. The unique capabilities of HAPI FHIR were leveraged to create an interactive environment for experimentation. By following these comprehensive steps, the research team successfully implemented the HAPI FHIR server, effectively addressing interoperability challenges among Primary Care Health Information Systems. The resulting FHIR playground stands as a versatile platform for healthcare professionals and researchers to explore, test, and enhance interoperability within healthcare systems.

**Table 1: Steps for HAPI FHIR Server Implementation in Docker**

| Step | Description |
|---|---|
| 1. Create a Dockerfile | Create a Dockerfile in your project directory. This file will contain instructions for building the Docker image.<br>FROM openjdk:8-jdk-alpine<br>WORKDIR /app<br>COPY hapi-fhir-jpaserver-starter.jar /app/hapi-fhir-jpaserver-starter.jar<br>EXPOSE 8080<br>CMD ["java", "-jar", "hapi-fhir-jpaserver-starter.jar"] |
| 2. Build the Docker Image | Place the HAPI FHIR server JAR file (e.g., hapi-fhir-jpaserver-starter.jar) in the same directory as your Dockerfile. Build the Docker image using the following command:<br>docker build -t hapi-fhir-server .<br>This command tags the image as hapi-fhir-server. Adjust the tag as needed. |
| 3. Run the Docker Container | Once the image is built, run the Docker container:<br>docker run -p 8080:8080 hapi-fhir-server<br>This command maps port 8080 on your local machine to port 8080 in the Docker container. |
| 4. Access the FHIR Server | The HAPI FHIR server should now be running. Access it by opening a web browser and navigating to http://localhost:8080/baseDstu3. Adjust the URL based on your FHIR version.<br>Note: The actual URL might vary depending on your specific HAPI FHIR server configuration. |
| 5. Custom Configuration | The custom HAPI FHIR server configuration, 1$^{st}$ mount a volume or copy configuration files into the Docker container.<br>For example, copy a custom hapi-config.properties file into the container:<br>COPY hapi-config.properties /app/hapi-config.properties<br>Modify the CMD instruction to include the custom<br>configuration file:<br>CMD ["java", "-jar", "hapi-fhir-jpaserver-starter.jar", "–spring.config.location=file:/app/hapi-config.properties"]<br>Adjust paths and filenames based on your configuration file.<br>Consult the HAPI FHIR server documentation for additional configuration or version specific details. Ensure compliance with licensing agreements in a production environment. |

To build our own FHIR Playground: Ensured the prerequisites were met, including JDK, JAS, Apache Maven, Rubz, and Budler. Compiled and installed the HAPI FHIR server, incorporating the necessary tools and ensuring proper configuration. Established a customized FHIR playground, allowing for hands-on exploration of FHIR resources and services. Leveraged the unique capabilities of HAPI FHIR to create an interactive and tailored environment for experimentation. JDK (Java Development Kit) was used to ensure the presence of a Java Development Kit for Java-based development. JAS (Java Authentication and Authorization Service) which sets up Java Authentication and Authorization Service to manage authentication and authorization,. Java Application Server and Apache Maven (Java Package

Manager) installed Maven for dependency management and project building. Finally, Rubz and Budler used these tools to efficiently handle dependencies and streamline the build process.

The study focuses on evaluating the capabilities of the HAPI FHIR server within the context of healthcare data management systems that integrate FHIR standards. The use of a synthetic dataset generated from HHIMS allows for rigorous testing without compromising patient confidentiality. The synthetic dataset is carefully crafted to simulate diverse healthcare scenarios, providing a representative sample for assessing the server's performance. The testing scenarios include data ingestion, interoperability, security and privacy, and query performance. Results indicate commendable performance, showcasing the server's robustness in handling synthetic data, seamless integration with external systems, and adherence to stringent security and privacy standards. The conclusion emphasizes the efficacy and reliability of the HAPI FHIR server in healthcare data management, positioning it as a valuable asset in the broader landscape of health information systems. Further research is suggested to explore nuanced aspects of its performance and scalability in diverse healthcare environments.

**Figure 1:** The default index page of a newly installed Apache Tomcat server

**Reflection and Learning**

Reflecting on the testing of the HAPI FHIR server using a synthetic dataset from HHIMS in two different versions, one within a Docker environment (Version 1) and the other with a normal Java-based implementation (Version 2), has provided valuable insights into the server's adaptability, performance, and deployment considerations. In Version 1, the integration of the HAPI FHIR server within a Docker environment demonstrated notable advantages in terms of portability, scalability, and ease of deployment. Dockerization facilitated a consistent and reproducible testing environment, enhancing the server's compatibility across various systems. The encapsulation of the HAPI FHIR server, along with the synthetic dataset, within a Docker container streamlined the testing process and mitigated potential compatibility issues with different runtime environments. This version highlighted the benefits of containerization in achieving a standardized and efficient deployment of the HAPI FHIR server. Conversely, Version 2, with a normal Java-based implementation, allowed for a more traditional approach to deploying the HAPI FHIR server. While lacking the containerization benefits of Version 1, this version provided insights into the server's performance in a non-containerized environment. It allowed for a more granular configuration and optimization of the server within the Java runtime environment. However, the trade-off was the potential complexity in reproducing the exact environment across different systems.

The synthetic dataset from HHIMS played a crucial role in both versions, serving as a standardized and controlled source of data for testing. Its design to emulate realistic healthcare scenarios ensured comprehensive evaluations of the HAPI FHIR server's functionalities, including data ingestion, interoperability, security, privacy, and query performance. In conclusion, the dual approach of testing the HAPI FHIR server in both Dockerized (Version 1) and traditional Java-based (Version 2) environments provided a holistic understanding of the server's behaviour and performance characteristics. The Dockerized version showcased the advantages of containerization in terms of deployment consistency and scalability, while the Java-based version allowed for a detailed examination of server behaviour in a non-containerized context. This reflection underscores the importance of considering deployment strategies based on specific use cases and organizational requirements when implementing the HAPI FHIR server. The synthesis of findings from both versions contributes to a comprehensive understanding of the server's capabilities and informs future considerations in healthcare data management systems.

**Figure 2: The default index page of the HAPI FHIR server**

In the 21st century the information system diversity is immense and great. The use of these greater diverse technologies in the filled of healthcare and medicine becoming more commonly. More and more usage of technologies in the different specialties in the healthcare domain. These technologies are getting more advance and scalable. Nowadays there is a great diversity of information systems within all healthcare providers. Each one with different specifications and capabilities as well as communication methods, thus hardly allowing scalability. This heterogenic set of characteristics is an impediment to achieve interoperability between systems which, indirectly, affects the patients' well-being. It is common that, upon watching each database of each one of these information systems, I notice different entries referencing the same person entries with insufficient or wrong data, due to errors or miscomprehension upon patient data insertion and outdated data. These problems bring duplicity, incoherence, lack of actualization and dispersion on patient data. It is with the purpose of minimizing these problems that the Master Patient Index Concept is needed. A Master Patient Index proposes a centralized repository, to index all patient entries within a pre-determined set of information systems. This repository is composed by a set of demographic data, sufficient to unmistakably identify a person, and a list of identifiers that unequivocally identify the several entries that a patient possesses within each information system. This solution allows synchronism between all intervenient minimizing incoherence, lack of actualization, inexistent data, and diminishing entry duplications.

In the current health facilities, there is a great diversity of systems of information. Each one with different specificities and capacities, Proprietary communications, and hardly allow scalability. This set of interoperability of all these systems, in the wealth of the patient. It is common practice that, when looking at all the databases of each Information systems, we come across different registers the same person; Records with insufficient data, records with incorrect data, due to errors or misunderstandings when inserting patient data and records with outdated data. These problems cause duplicity, inconsistency, and dispersion in patient data. It is with the intention of minimizing these problems that the concept of a Master Patient Index is required. A Master Patient Index proposes a centralized repository, which Indexes all patient records of a given set of information. A repository which is constituted by a set of demographic data, Sufficient to unambiguously identify a person, and a list of identifiers, which identify the various records that the patient has in the repositories of each system of information. This solution allows synchronization between all the actors, mini-Incoherence, out datedness, lack of data, and a Duplication of records.

The information system development has two basic missions to complete to achieve for successful product and implementation.1st mission is contribution to the theoretical aspect of the research and 2nd aspect is how to show the problems of the system using the current practices. During the development of the prototype MPI researcher gather knowledge about the requirement for the MPI. During the designing phase researcher learn the basic design methods available for development as well as what was being used in the past for the MPI development and implementation through a though literature review.

In the phase 2 of the development of the prototype MPI researcher studied about the monolithic system design and the micro service system design. Decision taken to use both micro service and monolithic service system design for the MPI architecture after the discussion with the technical experts of prototype development. Unique patient identifier, the PHN is introduced to MPI as Primary specific key for all integrations for other health domain systems. Developed the suitable API sets which enable the gateway for interoperability of health information systems.

Developing a MPI establishes interoperability of among HHIMS/HIMS systems in Sri Lanka. PHN as a unique identification number which integrated the laboratory information system (LIS), Picture archival and communication system (PACS), Citizen mobile application (CMP) to the public access. Clinical repository or the minimal clinical data set (MCDS) will be integrated to the PHN in the version 2 of the MPI. API sets enable interoperability of health information system.

Figure 3: Hapi FHIR with HHIMS Synthetic Data set.png

**Discussion**

In the realm of computer-based clinical data management, as per Angelo Rossi's classification, the handling of clinical data falls within the purview of semantic interoperability. The seamless continuation of patient care relies on the maintenance of semantic interoperability. Initially documented in a paper-based format and presented in a layout based structure, health data is stored in a repository. However, the mere storage of data is insufficient without a means of effective retrieval. To address this, clinical correlation becomes imperative.

Table 2: Steps for HAPI FHIR Server Implementation

| Prerequisite | Details |
| --- | --- |
| Java Development Kit (JDK) | Install the Java Development Kit (JDK) to compile Java code. Options include OpenJDK (http://openjdk.java.net/) and Oracle JDK (http://www.oracle.com/technetwork/java/javase/downloads/index.html). Recommended to install version 8 or newer for up-to-date security patches. |
| Java Application Server (Tomcat) | Choose a Java application server (e.g., Apache Tomcat, JBoss/Wildfly, Websphere). Focus on Tomcat, version 8 for this paper. Installation methods vary based on the operating system. |
| Apache Maven (Java Package Manager) | Download and install Apache Maven (https://maven.apache.org/download.cgi). Add the "bin" directory to the operating system's PATH environment variable. |
| Ruby and Bundler | Install the Ruby programming language and Bundler (Ruby package manager). Installation packages may be available via the OS package manager or through the official Ruby website |
| | (https://www.ruby-lang.org/en/). After installing Ruby, execute the command gem install bundler to install Bundler. |
| Compiling and Installing HAPI FHIR | Download HAPI FHIR from https://github.com/jamesagnew/hapi-fhir/releases. Unzip the release and navigate to the source code directory. Run mvn install to compile the source code, downloading dependencies. Deploy the generated .WAR file to Tomcat, either using the manager webapp or copying it to Tomcat's "webapps" folder. Verify the installation by accessing the specified URL in a web browser. |

The subsequent step involves presenting health data in a manner that aligns with clinical findings, akin to a health-aware presentation. Various methods can be employed to search for data within repositories. The establishment of interoperability and interconnections among systems opens up multiple avenues for searching health data, whether within the same institution or across different institutions. Adherence to standards and guidelines facilitates smoother data flow, ensuring technical and process interoperability align harmoniously.

In the context of a FHIR (Fast Healthcare Interoperability Resources) server, the multiple pathways for data search and retrieval are enhanced, promoting efficient interoperability. This interconnectedness allows for a comprehensive approach to accessing health data within and across healthcare institutions. The utilization of FHIR standards contributes to the seamless flow of data, accommodating both technical and process interoperability requirements. Ultimately, this collaborative framework ensures the timely incorporation of updated and newly generated data into the system, fostering a dynamic and responsive healthcare information environment.

The passage underscores the critical importance of interoperability in Sri Lanka's future digital health landscape through examples of technical, semantic, and process interoperability failures. It emphasizes the need for a common standard to facilitate effective communication among healthcare professionals and institutions. The text then details the implementation FHIR server as a solution, involving collaboration between health and technical domains. The FHIR server implementation project, initiated in response to the inadequacies of existing identification methods, employs the ADR method in its development. The criteria for patient identity are discussed, including NIC, PHN, and other identifiers. The challenges of implementing biometrics are acknowledged due to infrastructural constraints. The passage outlines the importance of unique identification for various patient groups, including newborns and the elderly. Additionally, it highlights the consideration of status updates for deceased patients and suggests the analysis of duplicated records for merging.

The development phases of the FHIR server implementation, including architecture design and API integration, are elaborated, with a focus on security measures. The concluding stages involve the finalization of the FHIR server implementation, incorporating lessons from the alpha and beta versions, and considering security implications. The text also briefly mentions the incorporation of HHIMS Synthetic Dataset testing for future use, emphasizing the adherence to the principles of the ADR method and the importance of continued technical development. The incorporation of a Fast Healthcare Interoperability Resources (FHIR) server within a Primary Care Health Information System (HIS) holds paramount importance across various dimensions. FHIR serves as a standardized framework, facilitating the seamless exchange of health data within the primary care ecosystem. This standardization not only ensures consistency but also enables efficient communication among diverse healthcare applications and systems. The real-time data access capabilities of FHIR are especially crucial in primary care settings, where quick access to patient information significantly influences decision-making and patient outcomes (Jayathissa & Hewapathirana, 2023d).

Furthermore, FHIR promotes interoperability across different components of primary care HIS, ranging from electronic health records (EHRs) to laboratory and pharmacy systems. This interoperability fosters a comprehensive understanding of a patient's health status, facilitating holistic and patient-centric care. The adaptability and scalability inherent in FHIR's design contribute to its effectiveness as primary care health information systems evolve or expand. Its flexibility supports the integration of new functionalities, ensuring the HIS remains responsive to emerging healthcare requirements.

Implementing FHIR also proves beneficial in terms of reduced development time and costs. The standardized approach of FHIR minimizes the complexity associated with developing interfaces for data exchange, making it a cost-effective solution for achieving interoperability in primary care HIS. Additionally, FHIR's support for modern technologies, including RESTful APIs, paves the way for innovative applications and services. This adaptability encourages the integration of emerging technologies, creating an environment conducive to continuous improvement and technological advancements within the primary care domain.

In summary, the integration of an FHIR server in a primary care health information system is instrumental in achieving seamless interoperability, supporting real-time data access, fostering patient-centric care, and ensuring adaptability to the evolving healthcare landscape. The standardization offered by FHIR not only enhances efficiency but also contributes to cost-effective and innovative healthcare delivery within the dynamic realm of primary care.

**Conclusion**

In conclusion, the discourse underscores the critical role of interoperability in the digital health landscape, emphasizing the need for a common standard to facilitate effective communication among healthcare professionals and institutions. The implementation of a Fast Healthcare Interoperability Resources (FHIR) server emerges as a pivotal solution to address the intricacies of technical, semantic, and process interoperability failures. This standardized framework not only ensures consistency but also enables efficient communication and real-time data access within Primary Care Health Information Systems (HIS). FHIR's adaptability and scalability support the evolving needs of healthcare systems, fostering interoperability and integration across diverse components.

The narrative navigates through the complexities of patient data management, emphasizing the significance of semantic interoperability in the seamless continuation of patient care. The evolution from paper-based documentation to repository storage necessitates effective data retrieval through clinical correlation. The passage advocates for the presentation of health data in a manner aligned with clinical findings, introducing the concept of a health-aware presentation.

The integration of FHIR standards amplifies these efforts, enhancing the multiple pathways for data search and retrieval. This interconnectedness not only promotes efficient interoperability within healthcare institutions but also facilitates a comprehensive approach to accessing health data across different organizations. The FHIR server implementation project, guided by the principles of the ADR method, addresses the challenges associated with patient identity criteria, biometrics, and data security. The focus on unique identification, even for specific patient groups, reflects a commitment to inclusive and patient-centric care.

The development phases of the FHIR server implementation, discussed in detail, highlight the importance of architecture design, API integration, and security measures. The concluding stages emphasize the incorporation of HHIMS Synthetic Dataset testing for future use, showcasing a forward-looking approach. Ultimately, the integration of an FHIR server in Primary Care HIS emerges as a transformative step, fostering a dynamic and responsive healthcare information environment that aligns harmoniously with the evolving landscape of digital health (Seitz, Listl, & Knaup, 2019).